\documentclass[iop,numberedappendix,revtex4]{emulateapj}
\usepackage{amsmath}
\usepackage{natbib}
\usepackage{threeparttable}

\newcommand{\be}{\begin{equation}}
\newcommand{\ee}{\end{equation}}
\newcommand{\ba}{\begin{eqnarray}}
\newcommand{\ea}{\end{eqnarray}}
\newcommand{\rhosph}{\rho^{\mbox{\tiny SPH}}}

\newcommand{\eg}{\textit{e.g.\ }}

\begin{document}
\title{Rapid Optimal SPH Particle Distributions in Spherical Geometries For Creating Astrophysical Initial Conditions}

\author{Cody Raskin\altaffilmark{1} \& J. Michael Owen\altaffilmark{1}}

\altaffiltext{1}{Lawrence Livermore National Laboratory, P.O. Box 808, L-038, Livermore, CA 94550}

\begin{abstract}

Creating spherical initial conditions in smoothed particle hydrodynamics simulations that are spherically conformal is a difficult task. Here, we describe two algorithmic methods for evenly 
distributing points on surfaces, that when paired can be used to build 3D spherical objects with optimal equipartition of volume between particles, commensurate with an arbitrary, radial 
density function. We demonstrate the efficacy of our method against stretched lattice arrangements on the metrics of hydrodynamic stability, spherical conformity, and the harmonic power 
distribution of gravitational settling oscillations. We further demonstrate how our method is highly optimized for simulating multi-material spheres, such as planets with core-mantle
boundaries. 

\end{abstract}

\section{Introduction}

Smoothed Particle Hydrodynamics (SPH) \citep{gingold1977,lucy1977} is a meshless hydrodynamics scheme widely used in many astrophysical contexts. Being Lagrangian, SPH does 
not require regularly spaced, lattice-oriented interpolants as Eulerian mesh methods do. However, there is currently a paucity of readily available initial conditions generators for 
Lagrangian particle distributions. As a result many SPH users resort to using latticelike particle arrangements, and where sphericity is required they often simply apply a clipping operator 
to these regular arrangements to produce a subset of particles that occupy a sphere (\eg \cite{benz1987}, \cite{monaghan1991}, \cite{bate1998}, \cite{rasio1999}). These latticelike 
arrangements may be cartesian cubical lattice (CL), a hexagonal close packing (HCP) arrangement (\eg \cite{kitsionas2002}, \cite{davies1991}, \cite{davies1992}), or a quaquaversal 
tiling (QVT) arrangement \citep{hansen2007}, each of which use some space-filling method to achieve equipartition of volume.
For brevity we will refer to all such permutations of regular space-filling methods that are not necessarily spherically conformal as ``lattices''.

For many spherical problems, especially those involving shocks, lattice particle arrangements introduce a host of distorting features into SPH simulations. For example, shock-driven, 
radially compressive flows may demonstrate imprinting along one or more of the regular cardinal directions inherent in the underlying latticelike point arrangement as the particle column 
density is higher along those directions \citep{herant1994}. More importantly, applying an arbitrary radial profile to particles arrayed in a lattice is difficult under the constraint of equal mass 
particles throughout the distribution. \cite{rosswog2008} adapted a method of stretching a uniform CL in the radial direction so as to reproduce an arbitrary density profile, however, this 
method does not obviate the drawbacks of a lattice as we discuss below.

There are several alternative methods to lattice distributions for the purpose of generating spherical distributions of points, and these fall into two major categories -- one-stage and two-stage 
setups. Typical one-stage setups rely on random distributions according to some arbitrary probability distribution. The simplest of these is the Monte Carlo method, wherein the entire 
sphere is populated randomly by picking numbers from a probability distribution that matches the desired radial profile. A slightly more sophisticated approach populates shells at 
predetermined radii with randomly selected locations in the chosen shell. For the purposes of this paper, we will refer to the latter approach as the Random Shell method.

Two-stage setups take the output of any of the previously described methods (including lattice arrangements) and apply some extra iterative physics on the particle positions to drive 
toward the goal of nearly equal volume or surface area per point. One such approach is the Gravitational Glass (GG) method of \cite{wang2007} which uses an inverse gravitational field 
in conjunction with a motion-dampening force to drive particles toward an optimally spaced configuration. When combined with a lattice stretching algorithm, this method can also 
reproduce density profiles well, depending on the application. Another alternative is the Concentrated Shell setup, which uses the Random Shell method as its starting point. In this 
method, the particles are given a repulsive force from one another while constrained to remain inside their shells, which also settles them into a nearly optimally spaced distribution within 
each shell \citep{fryer2002,hungerford2003,fryer2007late}.

\cite{diehl2012} have developed a more sophisticated alternative, two-step method using a GG-type repulsion force inspired by weighted Voronoi tessellations (WVT). This 
method harnesses the power of an oct-tree to optimally arrange particles in three dimensions via a method similar to, but subtly different from the Lloyd algorithm \citep{lloyd1982}. 
While this method produces a reasonably relaxed distribution, it can be computationally expensive and does not entirely eliminate local shot noise. For their purposes, the authors 
of WVT use a Monte Carlo method for the first stage.

Each of these two-stage methods produces particle distributions near the low-energy, optimal configuration with varying success. However, each of 
them also requires a complicated and potentially expensive setup routine in combination with a relaxation step, especially where GG-type repulsion forces are required. For 
the most part these methods require a separate physics code for the sole purpose of initial conditions. While accurate and high-fidelity initial conditions are often crucial 
to getting a trustworthy numerical result, occasionally computational expense and time expediency argue for something simpler, yet still robust. In this paper, we describe a 
computationally simple, one-stage method for optimal or near-optimal spherically conformal arrangements that is rapid, easy to deploy for any SPH code, and requires no tree knowledge or expensive 
repulsion/relaxation pre-step. In Section 2, we describe our method which employs a hybrid of two different methods for arranging particles into shells, and in Section 3, we 
compare the results of our particle arrangements to a stretched CL arrangement (the most commonly used \textit{just in time} alternative), a random shell method (where particles
are randomly distributed in shells), and a Monte Carlo method (particles are randomly distributed in all three dimensions according to a probability distribution) on a variety of metrics 
during the creation and hydrodynamic relaxation of a two-material, Earth-like object. In Section 4, we give our conclusions.

\section{Method}

Distributing an arbitrary number of equal mass points inside a sphere according to a density function is not a simple task. We can simplify the problem by reducing our three degrees of 
freedom to two if we parameterize the radial coordinate according to our density function such that we are left with the problem of building successive shells of particles. At each radial 
coordinate ($r_i$), the total number of particles in the shell is
\be
 N=\frac{4\pi}{m_0}\int_{r_{i-1}}^{r_i} \rho(r) r^2 dr,\label{eq:shell} 
\ee 
where $m_0$ is the nominal mass per node. What we need now is a method of evenly distributing $N$ particles on a spherical surface. Our method employs two nested, algorithmic distribution schemes for populating shells in a hybrid fashion.

\subsection{RPR scheme}

\begin{figure}[ht]
\centering
\includegraphics[width=0.45\textwidth]{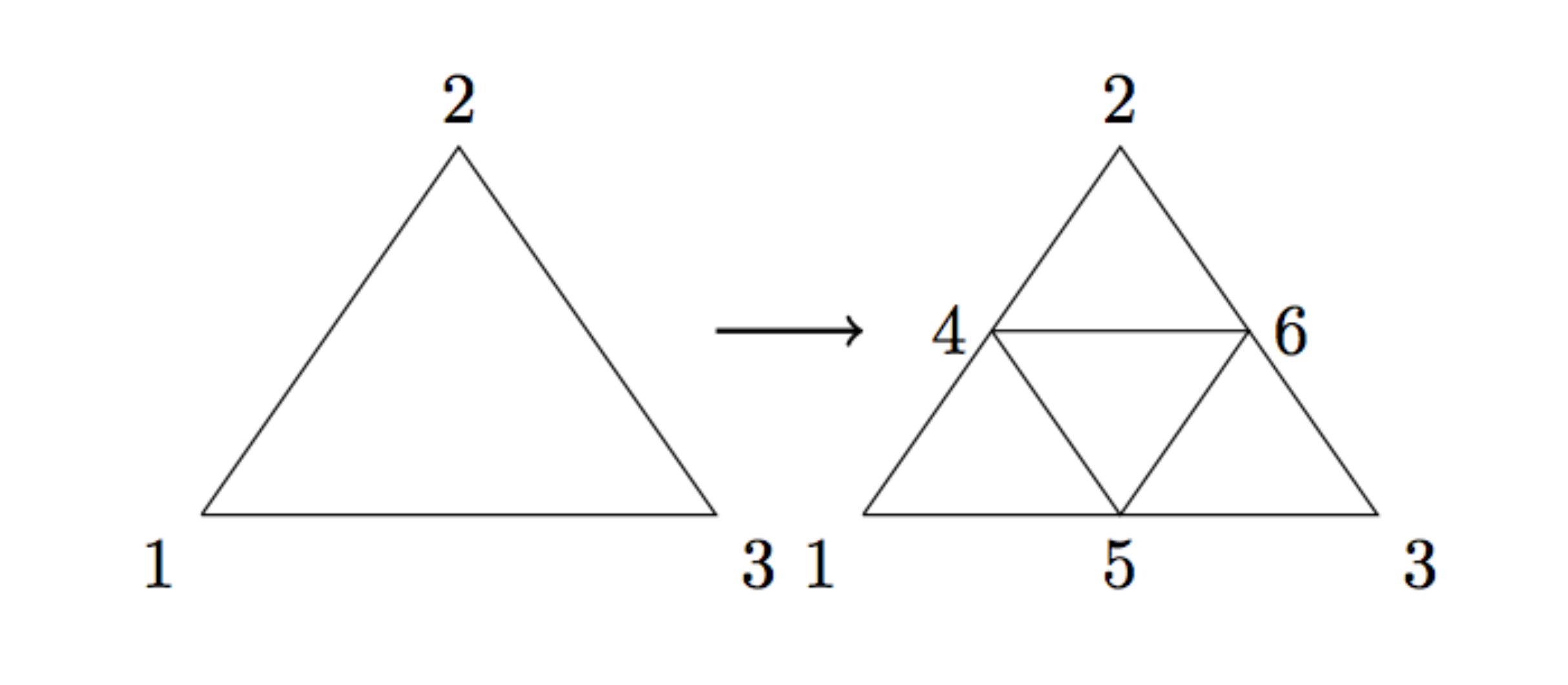}
\caption{A graphical representation a single-level triangle refinement.}
\label{fig:triangle}
\end{figure}

The first half of the method relies on a recursive primitive refinement (RPR) algorithm. The primitives are three dimensional solids with triangular facets (Platonic shapes), 
and we refine on those triangles to create shells of particles at successively higher particle counts. This procedure was described by \cite{herant1994}. The goal is to 
maintain equal mass particles in each shell and between shells with roughly equal spacing in all directions between particles and their local neighbor set. We accomplish 
this via a Catmull-Clark subdivision process \citep{catmull1978}. As shown in Figure \ref{fig:triangle}, a single refinement ($n=1$) of one triangle produces four similar 
triangles by bisecting the $n-1$ triangle edges at their midpoints. This produces three new vertices (4,5,6) that will each be one vertex of a refinement of an adjacent 
triangle. 

The storage of these new vertices may be handled abstractly (so as not to double-count vertices created through edge bisections), and the particle positions for these vertices are displaced by unit vectors scaled to an appropriate radial displacement after the recursive refinement process has completed. This process can be performed on any primitive polygon built of triangles, such as a tetrahedron or icosahedron.

\begin{figure}[ht]
\centering
\includegraphics[width=0.23\textwidth]{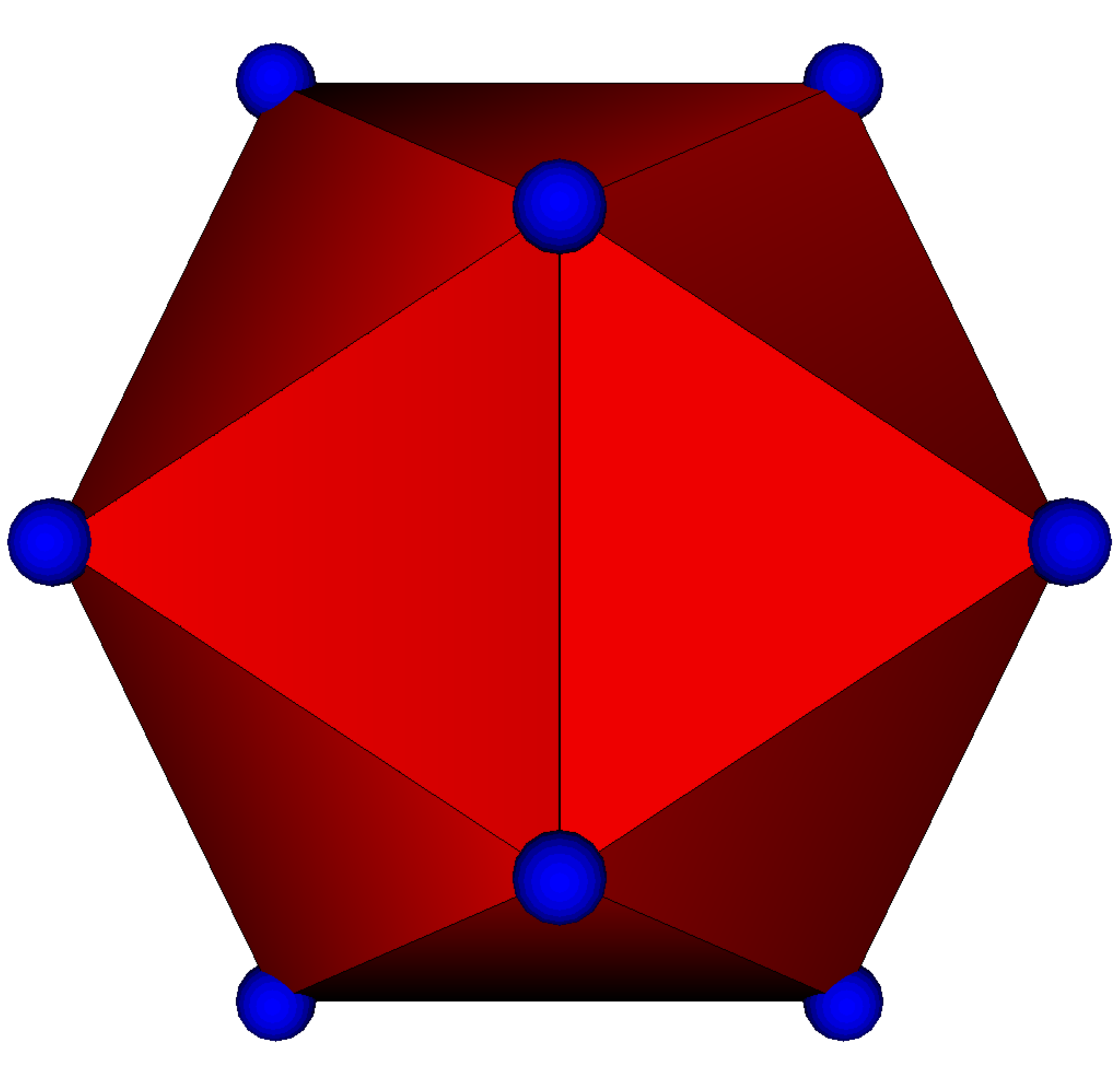}

\includegraphics[width=0.23\textwidth]{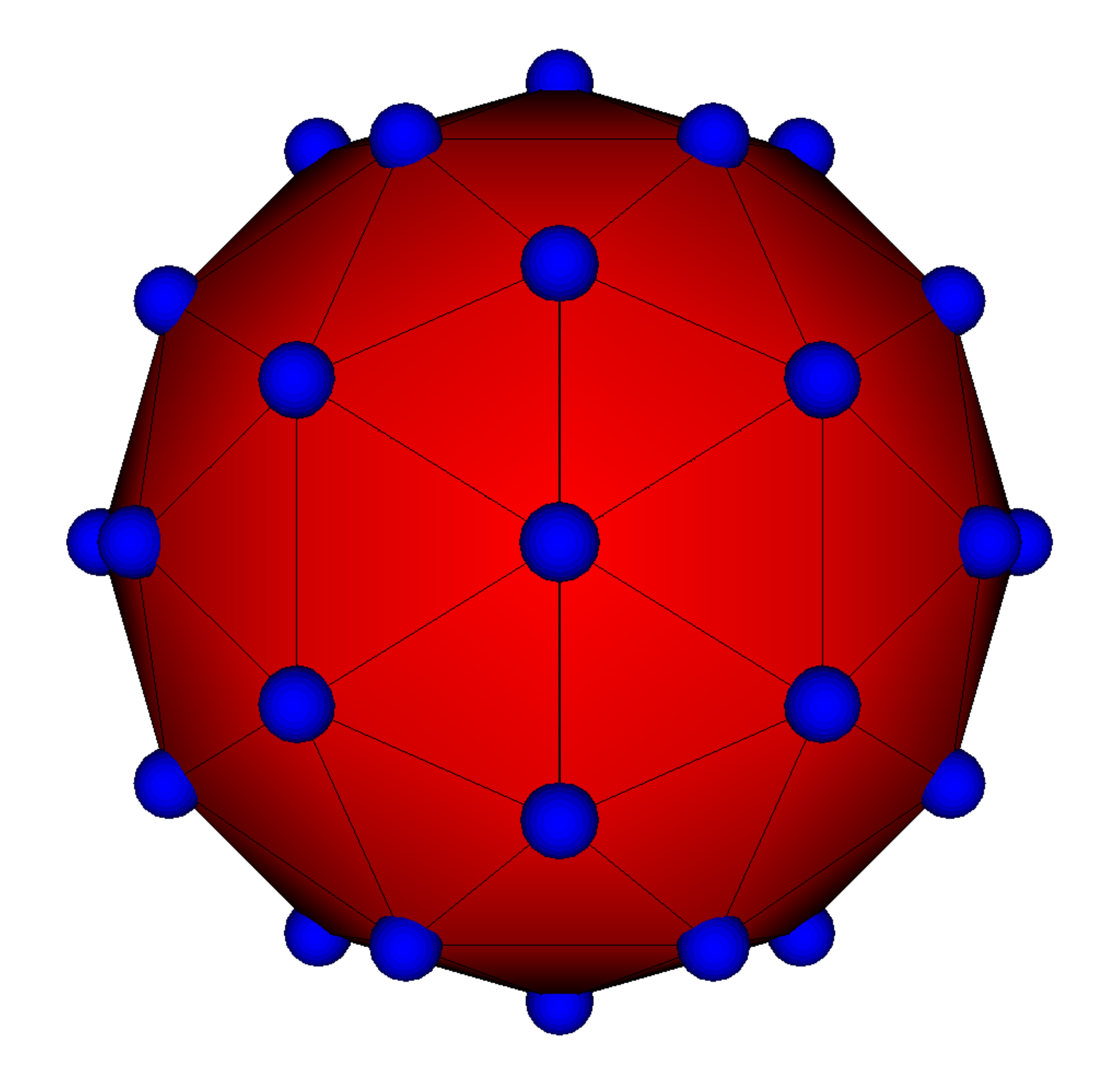}

\includegraphics[width=0.23\textwidth]{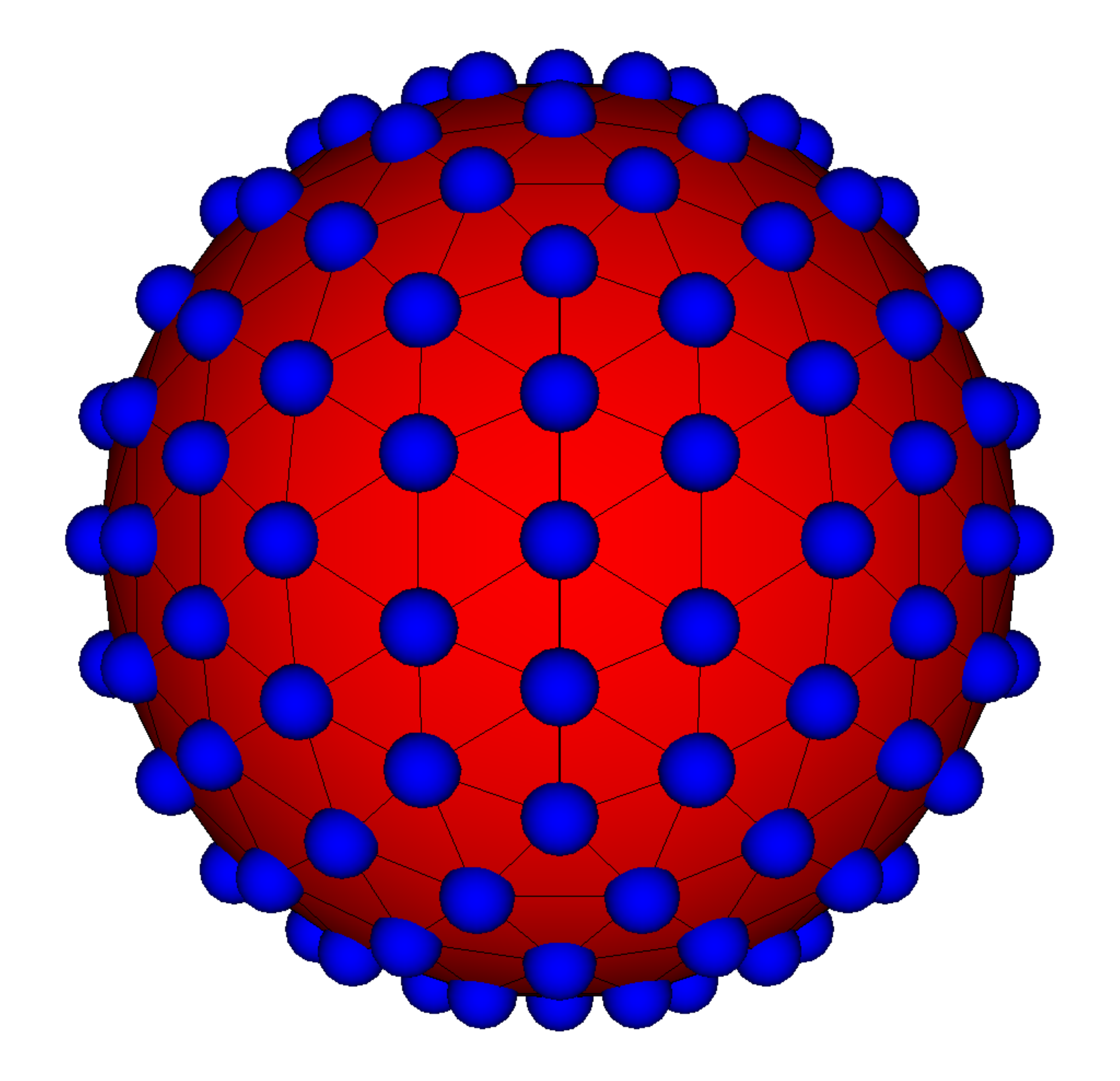}
\caption{Icosahedra with $n=0$ refinement (12 points), $n=1$ (42 points), and $n=2$ (162 points), respectively.}
\label{fig:icosahedron}
\end{figure}

For the purposes of creating nested spherical shells with equal mass particles, we first compute the total integrated mass for a given radial density distribution. A user may desire an 
arbitrary radial particle resolution, and so this fixes a nominal particle mass for the entire distribution. Each successive radial shell particle count can be computed simply by the shell 
mass divided by the nominal particle mass (as in equation (\ref{eq:shell})). At this stage, one might desire a sufficiently large number of primitive types so as to fill in any gaps between 
successive refinement levels. Consider for instance refining icosahedra, as is done in Figure \ref{fig:icosahedron}. At $n=3$ refinement levels, the total number of surface points is already 
642 -- the next lowest shell count using only icosahedra would be 162 at $n=2$. If, in order to maintain equal masses between shells, one desires $\approx 300$ particles in a shell, there 
exists no refinement level of an icosahedron that will produce particles within $\sim 2 \times$ the desired, nominal particle mass. These gaps between successive shell counts for a given 
primitive shape grow exponentially larger at high refinement levels ($\sim 2^{2n-1}$).

\begin{figure*}[ht]
\centering
\includegraphics[width=0.20\textwidth]{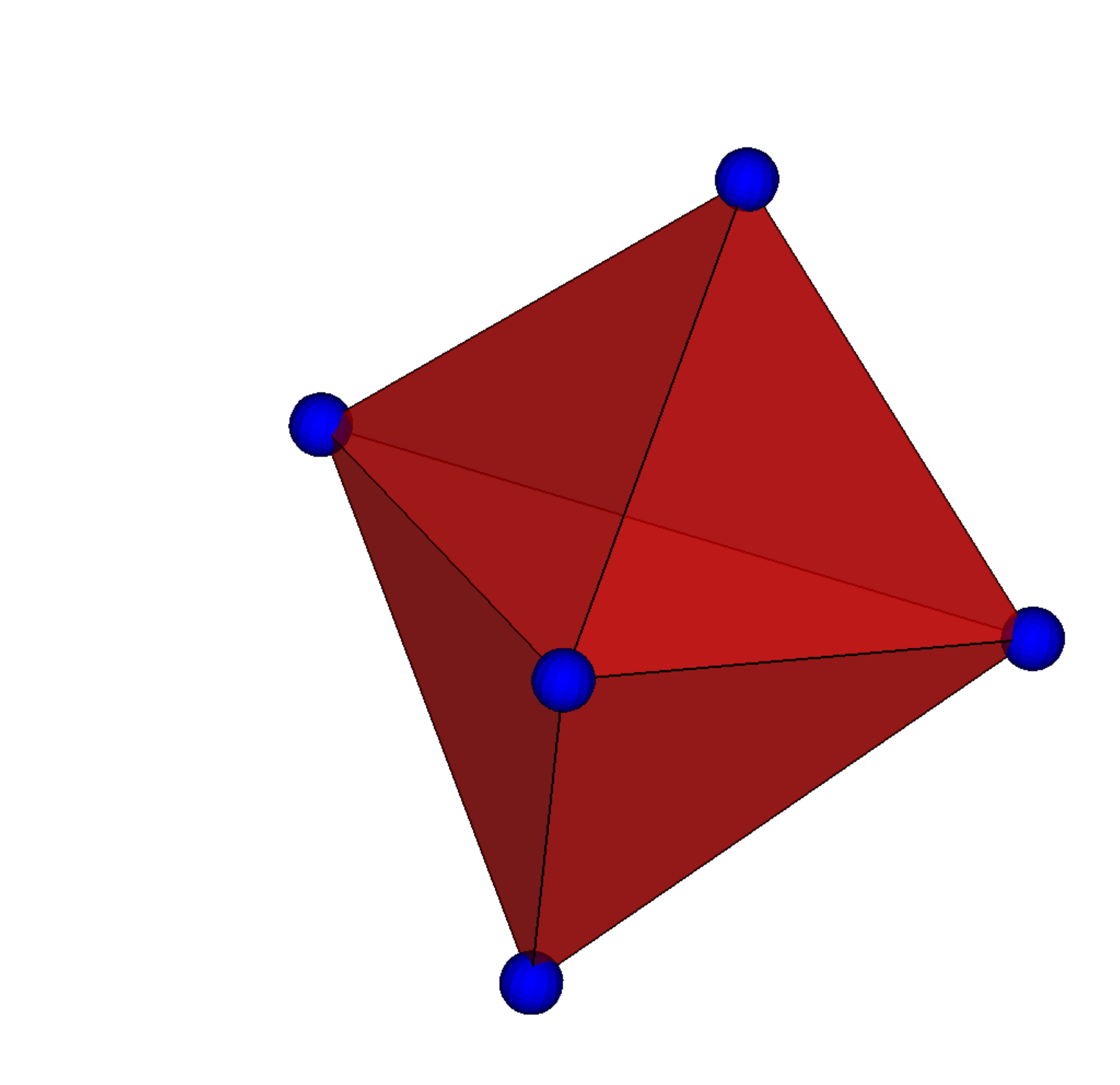}
\includegraphics[width=0.20\textwidth]{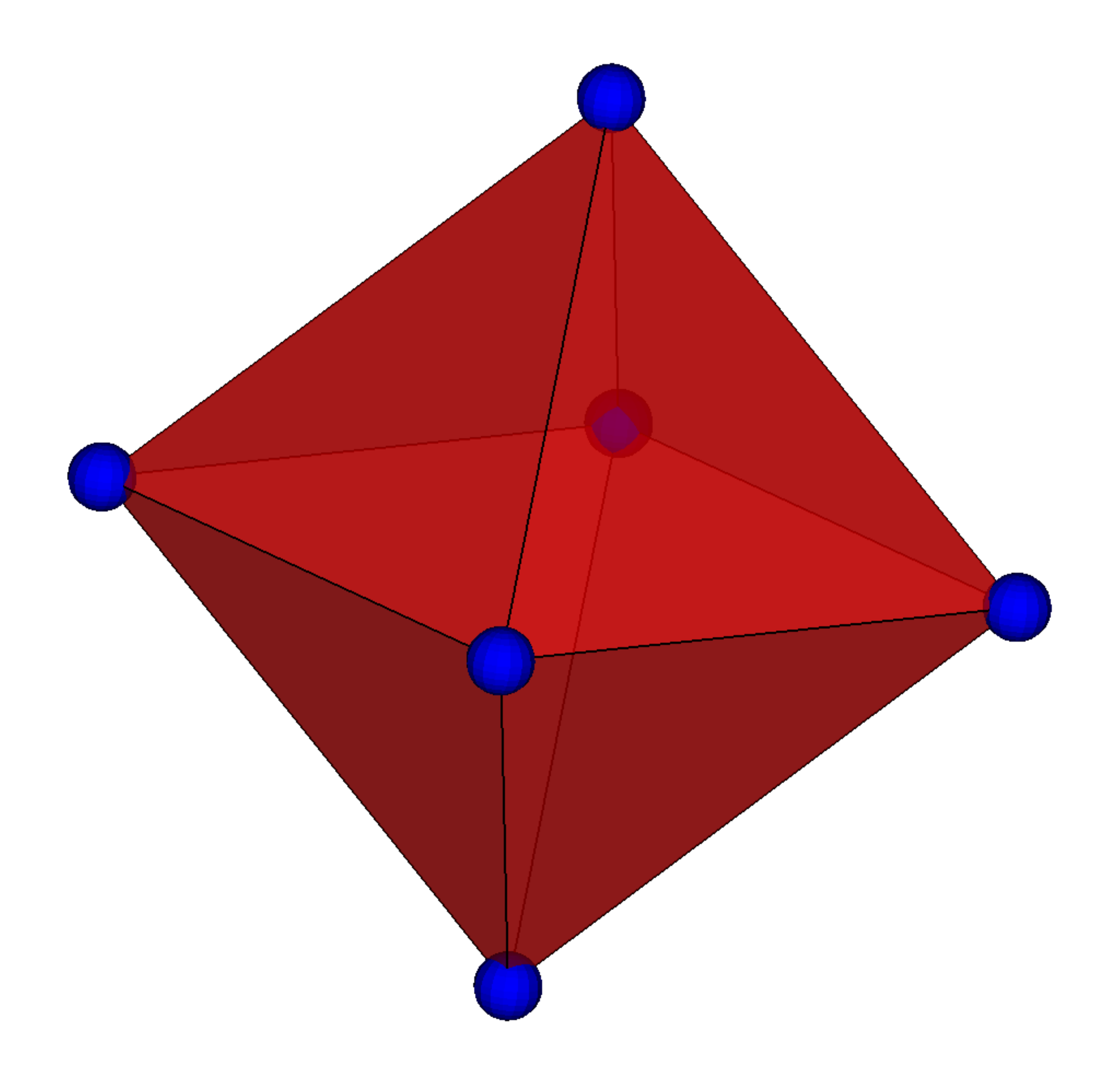}
\includegraphics[width=0.20\textwidth]{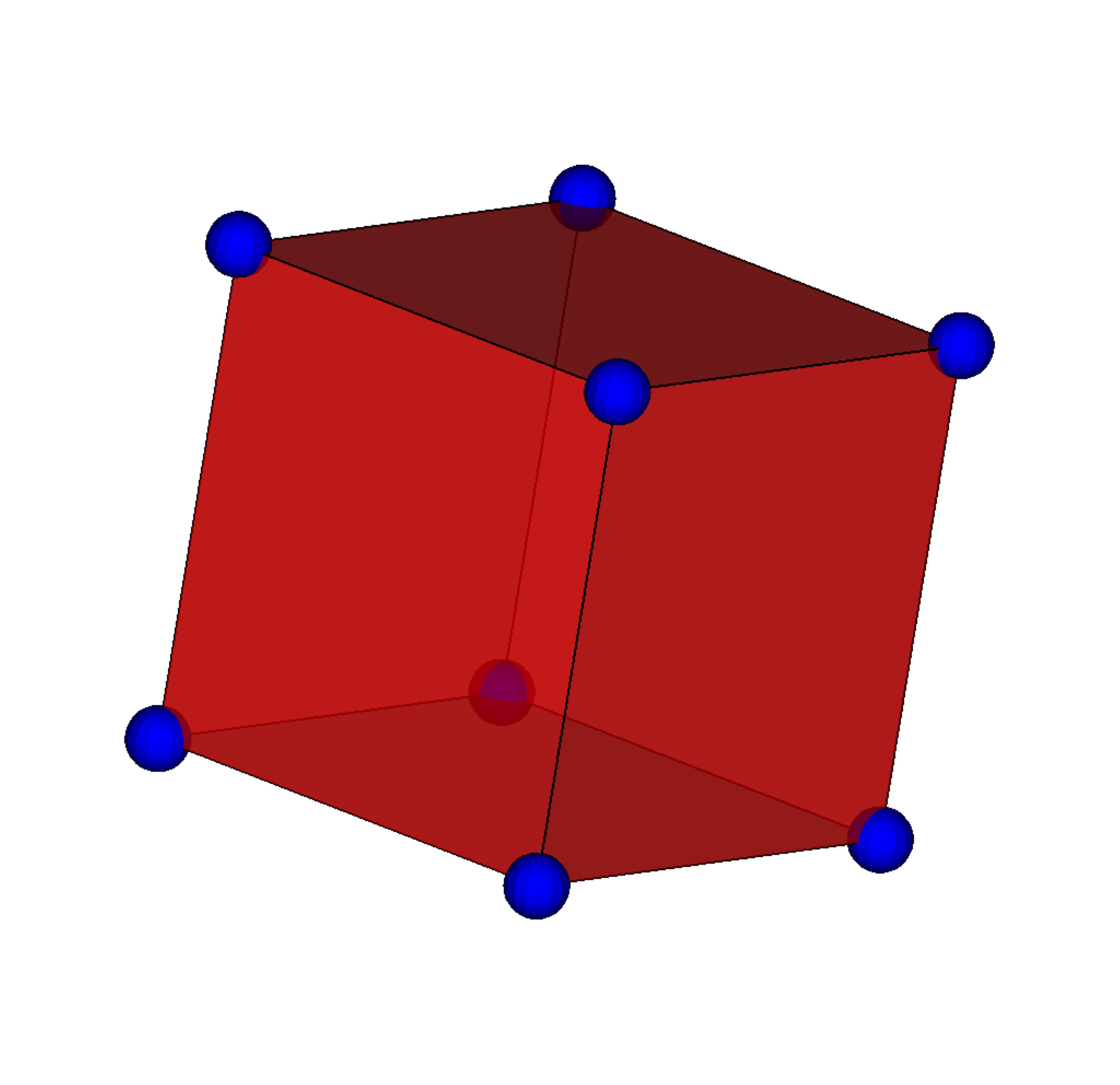}
\includegraphics[width=0.20\textwidth]{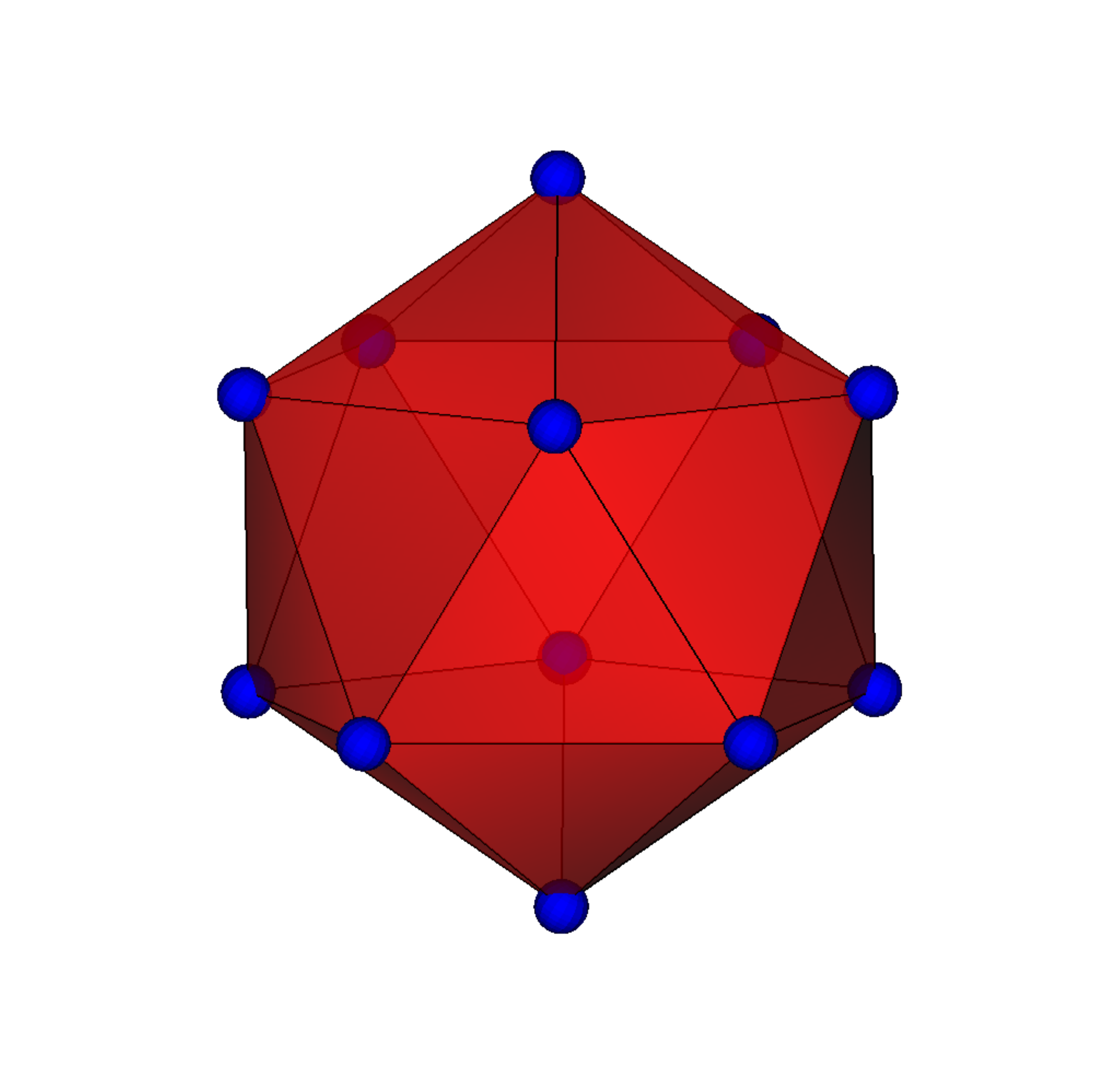}
\caption{The four primitive shapes (in this case, the first four Platonic solids) we use in conjunction with recursive refinement to build particle shells at desired particle counts. 
In the case of the cube, any two diagonal points on a face can be chosen arbitrarily to form an edge of a triangle such that the primitive $n=0$ shape has 12 faces.}
\label{fig:library}
\end{figure*}

To combat this problem, we store a library of primitive types (depicted in Figure \ref{fig:library}) whose refinement particle counts neatly span the counts between successive refinements of the various other 
primitive types. In order to choose the correct primitive type for a desired shell count, the total number of points on the surface of a primitive shape that has been refined $n$ 
times can be calculated by
\be
N_p(n) = N_\Delta N_t(n) - N_e N_{pe}(n) + N_c,
\ee
where $N_\Delta$ is the number of triangles in the unrefined primitive shape, $N_t(n)$ is the number of points in a triangle that has been refined $n$ times, given by
\be
N_t(n) = 2^{(2n-1)} + 3\times2^{(n-1)} + 1,
\ee
$N_e$ is the number of edges of the primitive shape, $N_{pe}(n)$ is the number of points along an edge of a primitive shape that has been refined $n$ times, given by
\be
N_{pe}(n) = 2^n + 1,
\ee
and $N_c$ is the number of corners of the primitive shape. Using this formula, we bracket the desired shell particle count between two different shapes at arbitrary refinement levels, 
and we choose the closest shape refinement to the desired count. In practice, this results in particle masses that vary by roughly $\sim1.2\times$ at most, and adjacent shell particle 
mass ratios are near unity. Figure \ref{fig:library} shows the primitive shapes we employ for our RPR scheme that, together with recursive refinement, can be used to construct shells 
of a variety of particle counts. For shells near the center of the sphere (where particle counts are near unity) we simply choose which is closest to the desired particle mass of either 
a one, two, or four particle arrangement about the center of the system -- the four particle arrangement being that of a regular tetrahedron.

\subsection{PS scheme}

The RPR scheme works very well for low to mid-range shell particle counts, but as the shell counts grow large, the unequal area mapping in the projection of the primitive triangular 
shapes onto a spherical surface introduces artifacts in the final arrangement. To mitigate this problem, we restrict the use of the RPR scheme to a maximum shell particle count of 
162 (icosahedron $n=2$). For shells with higher particle counts, we employ the parameterized spiraling (PS) scheme described by \cite{saff1997}.

\begin{figure}[ht]
\centering
\includegraphics[width=0.35\textwidth]{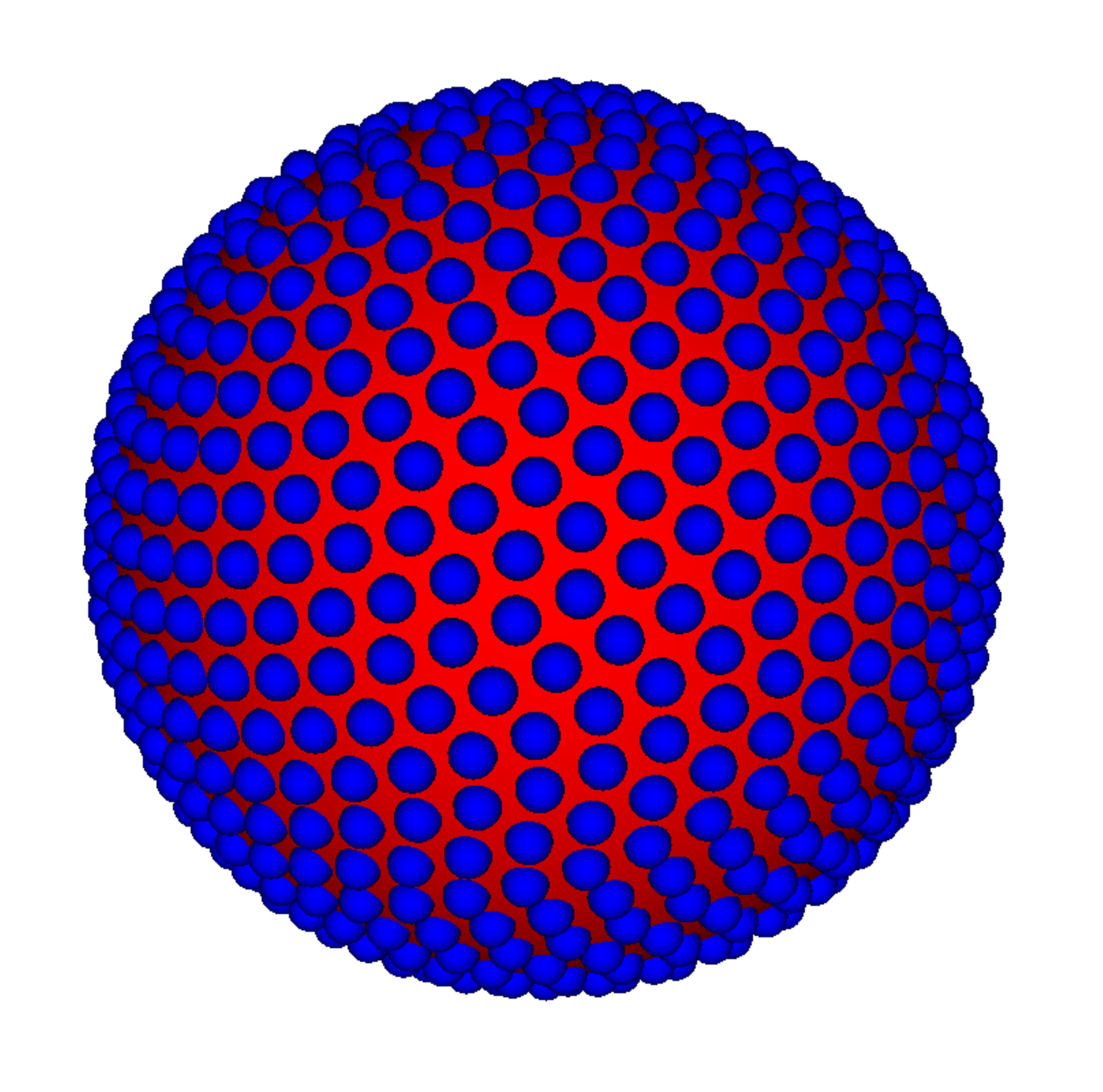}
\caption{A PS distribution for a shell count of 800 particles.}
\label{fig:spiral}
\end{figure}

In the PS scheme, the two principle angles in spherical coordinates are parameterized with a stepping parameter
\ba
h_k=-1+\frac{2(k+1)}{(N-1)},&1\leq k \leq N,
\ea
such that
\ba
\theta_k &=& cos^{-1}(h_k),\\
\phi_k &=& \left(\phi_{k-1}+\frac{3.8}{\sqrt{N}}\frac{1}{\sqrt{1-h_k^2}}\right),
\ea
where $N$ is the total number of particles in the shell and $\phi_0=\phi_N=0$. In our notation, $\theta$ is the polar angle. The choice of 3.8 in the definition of $\phi_k$ comes from the close packing argument laid out 
in \cite{vanderwaerden1968}, namely that $(\phi_k-\phi_{k-1})\sqrt{1-h_k^2}\approx 3.8/\sqrt{N}$. Figure \ref{fig:spiral} demonstrates the ability of the PS scheme to 
efficiently distribute large numbers of particles on a spherical surface.

The PS scheme is quite adept for large shell particle counts, but falls short for low shell particle counts as the stepping in $\phi$ becomes chaotic and poorly sampled as 
$N\to1$. Therefore, it is ideal to combine the RPR and PS schemes at low and high shell particle counts, respectively.

Finally, to avoid producing columns of particles along each of the poles we rotate each shell from both the PS and RPR distributions in $\phi$ and $\theta$ with random numbers seeded from the shell particle 
count. This way, the final arrangements do not have any preferred degrees of freedom and the random rotations are reproducible for a given shell count.

\begin{figure*}[ht]
\centering
Ia)\includegraphics[width=0.20\textwidth]{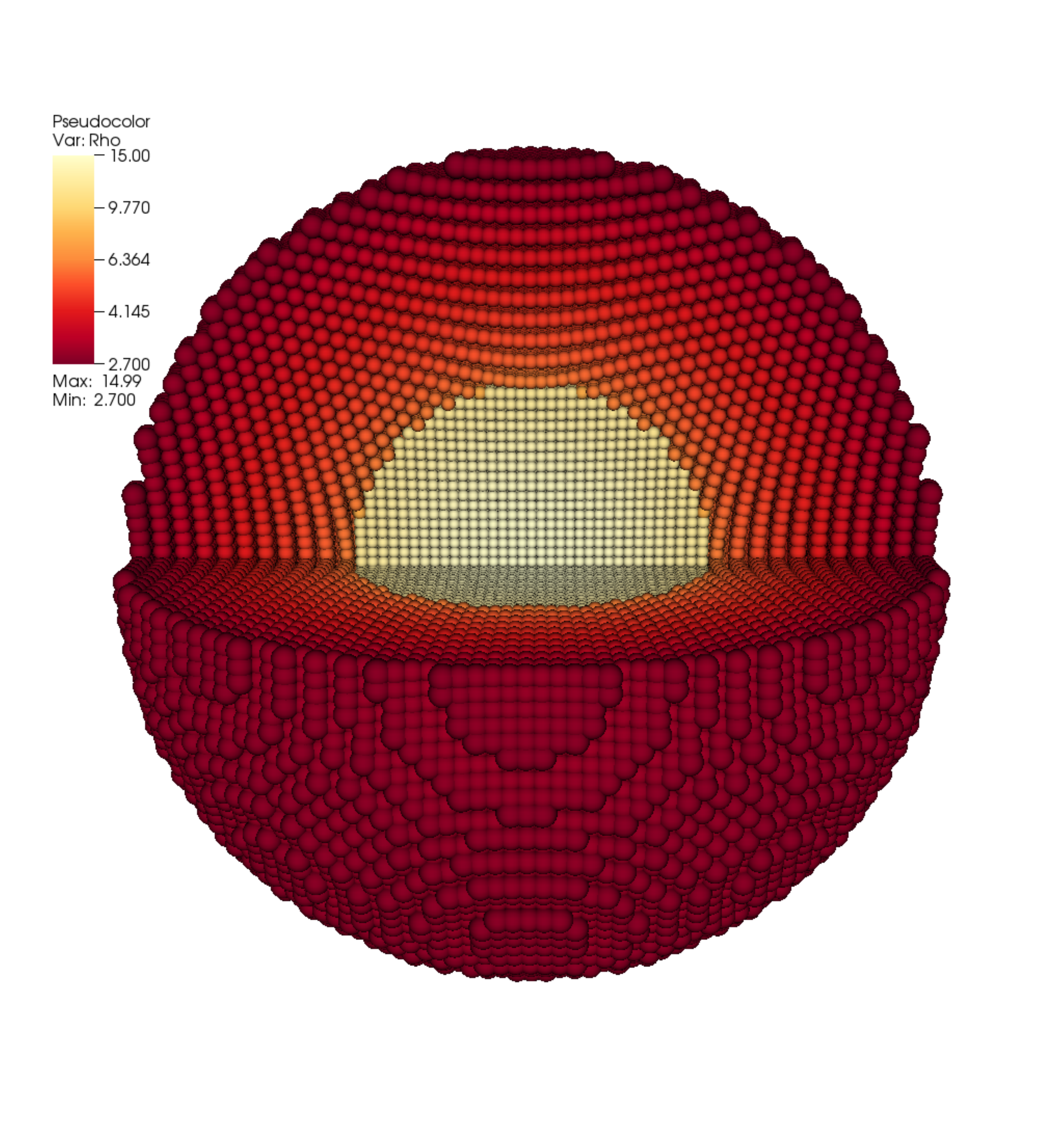}
Ib)\includegraphics[width=0.20\textwidth]{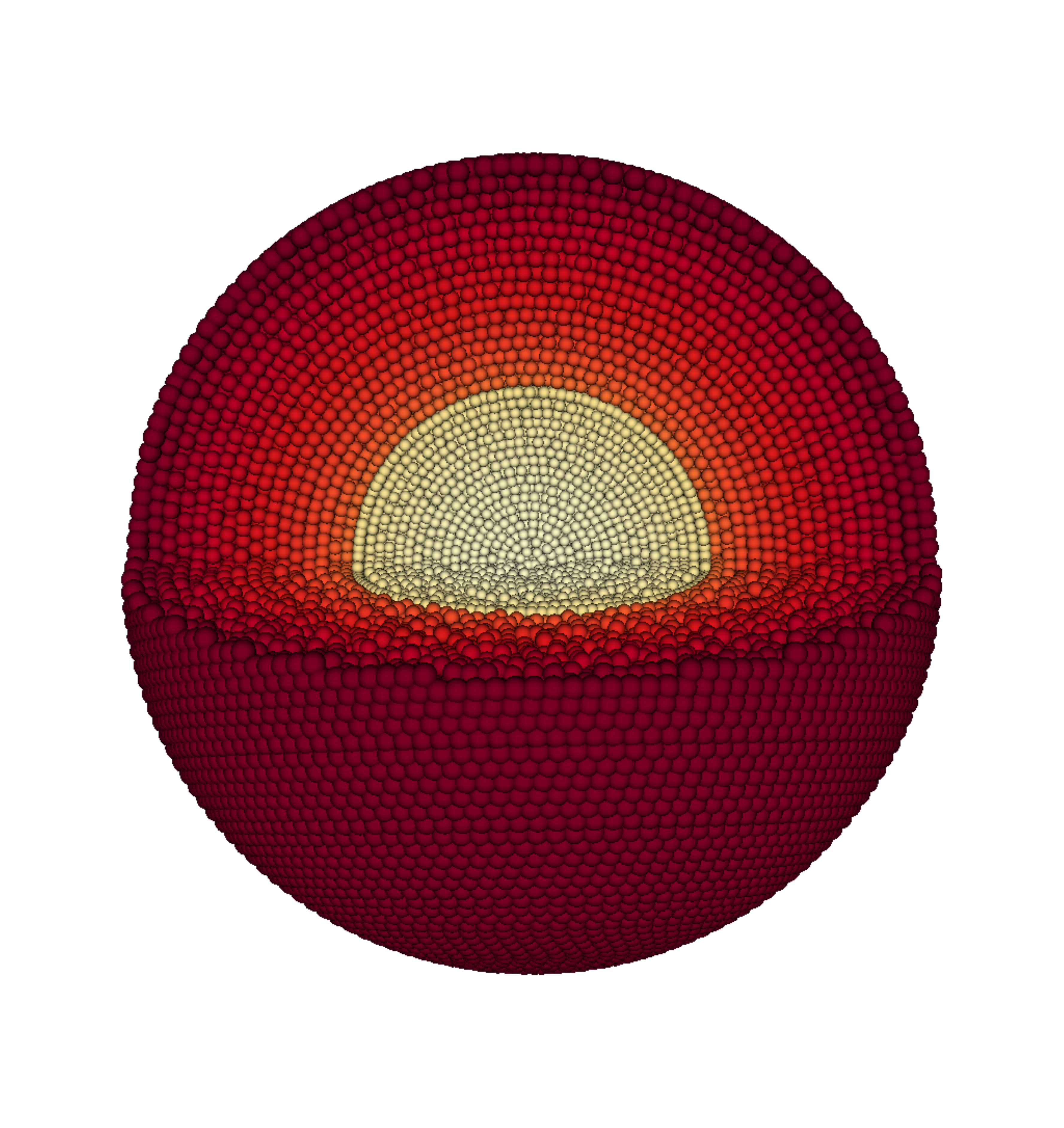}
Ic)\includegraphics[width=0.20\textwidth]{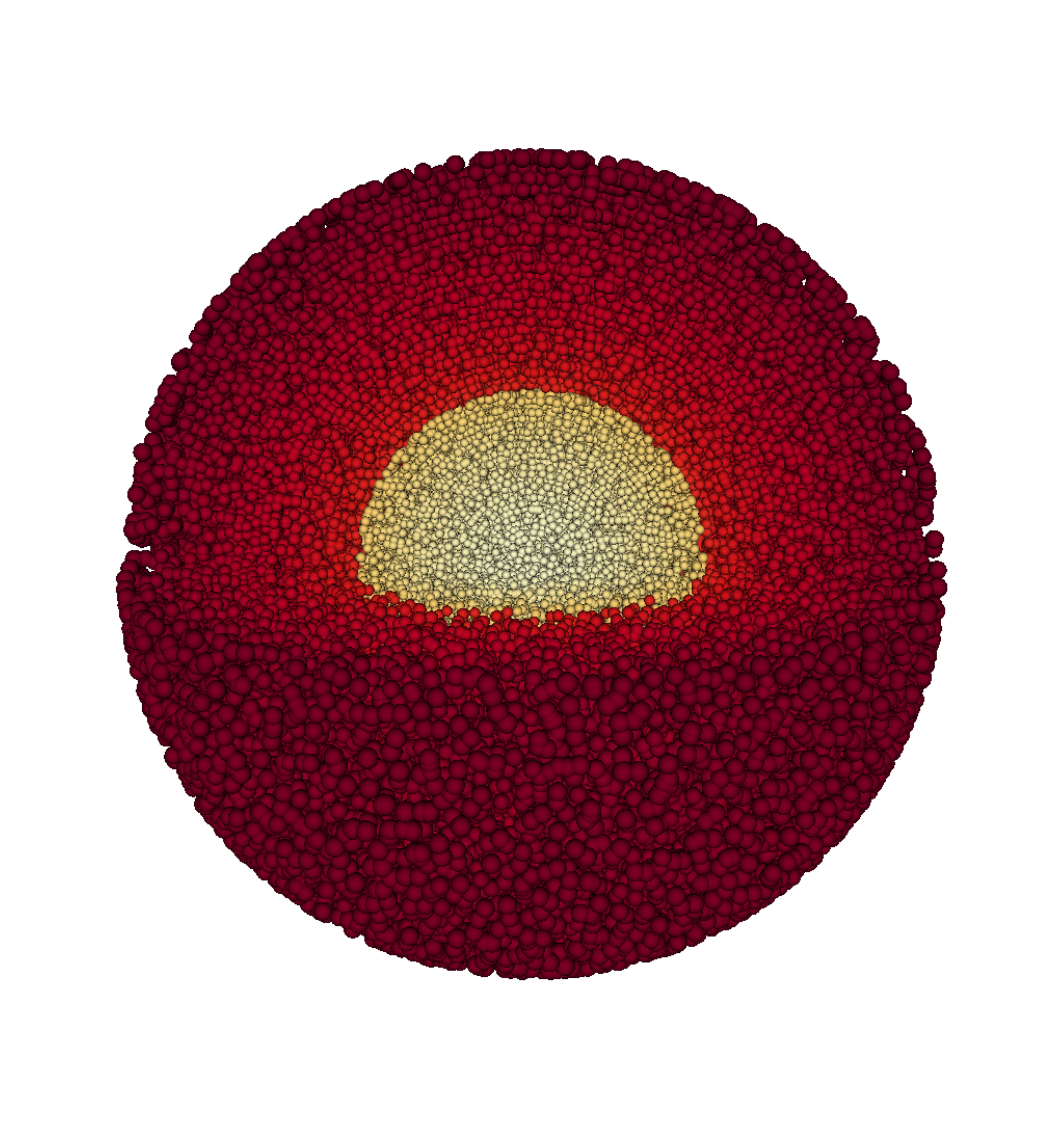}
Id)\includegraphics[width=0.20\textwidth]{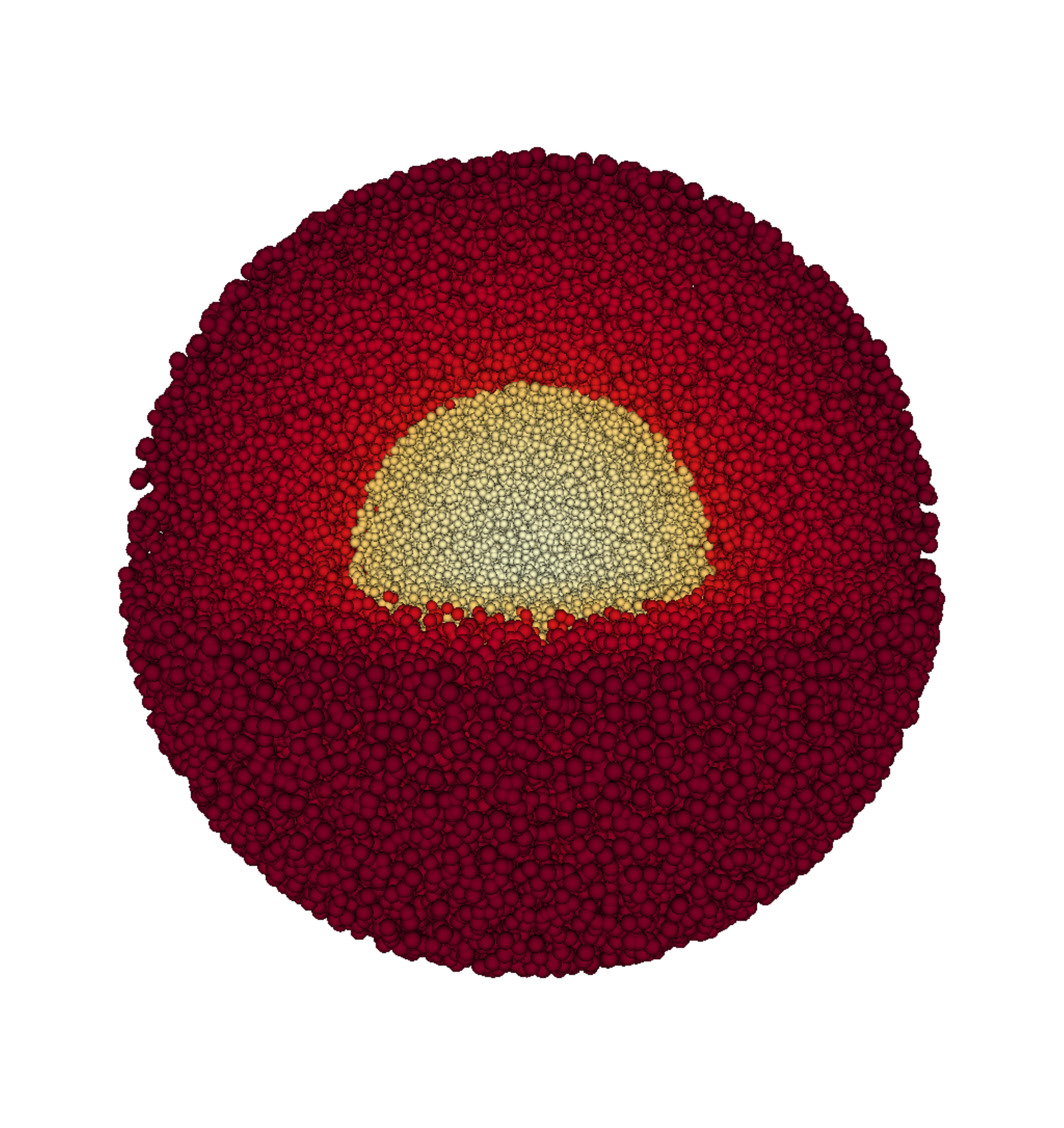}\\
IIa)\includegraphics[width=0.20\textwidth]{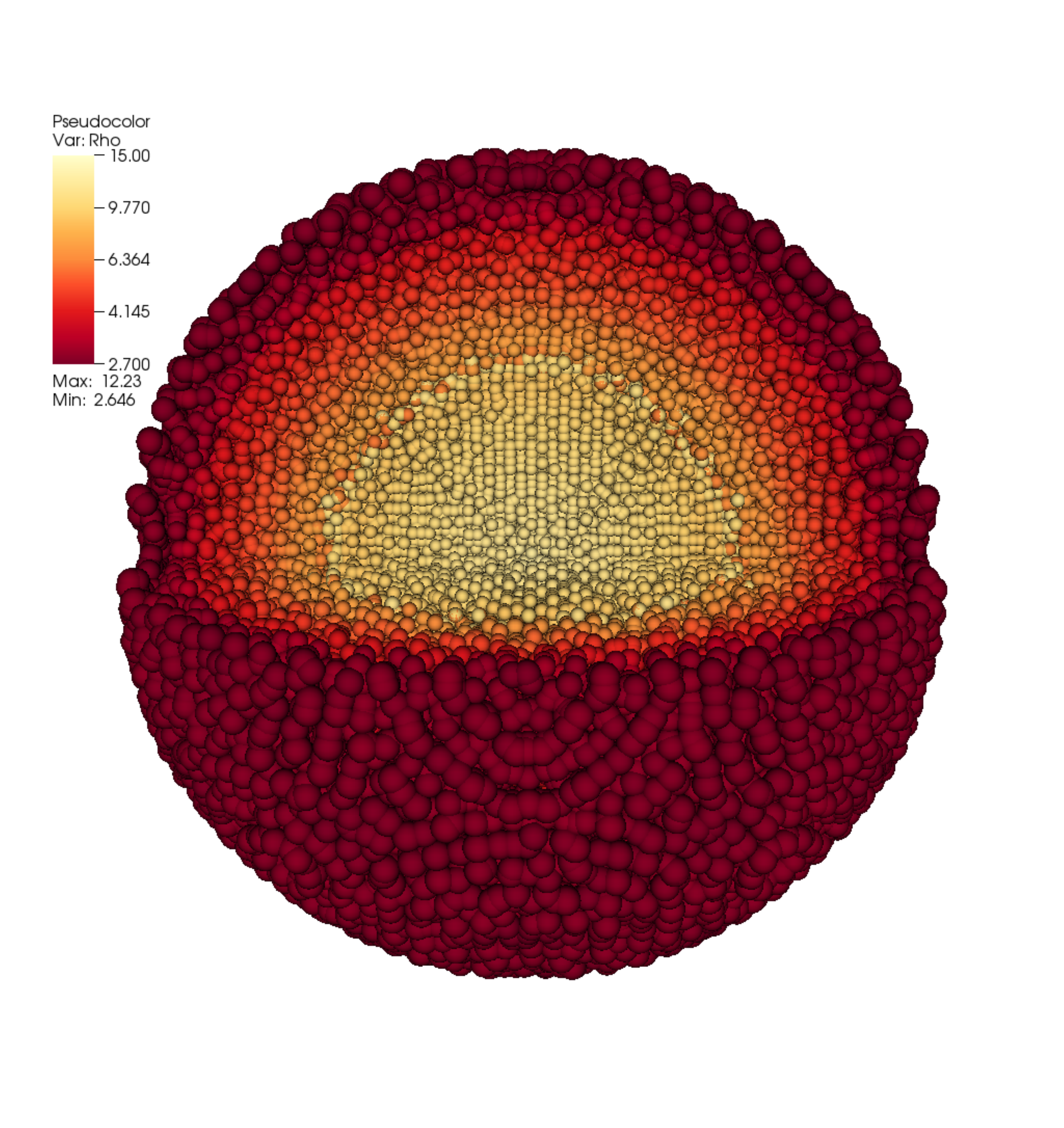}
IIb)\includegraphics[width=0.20\textwidth]{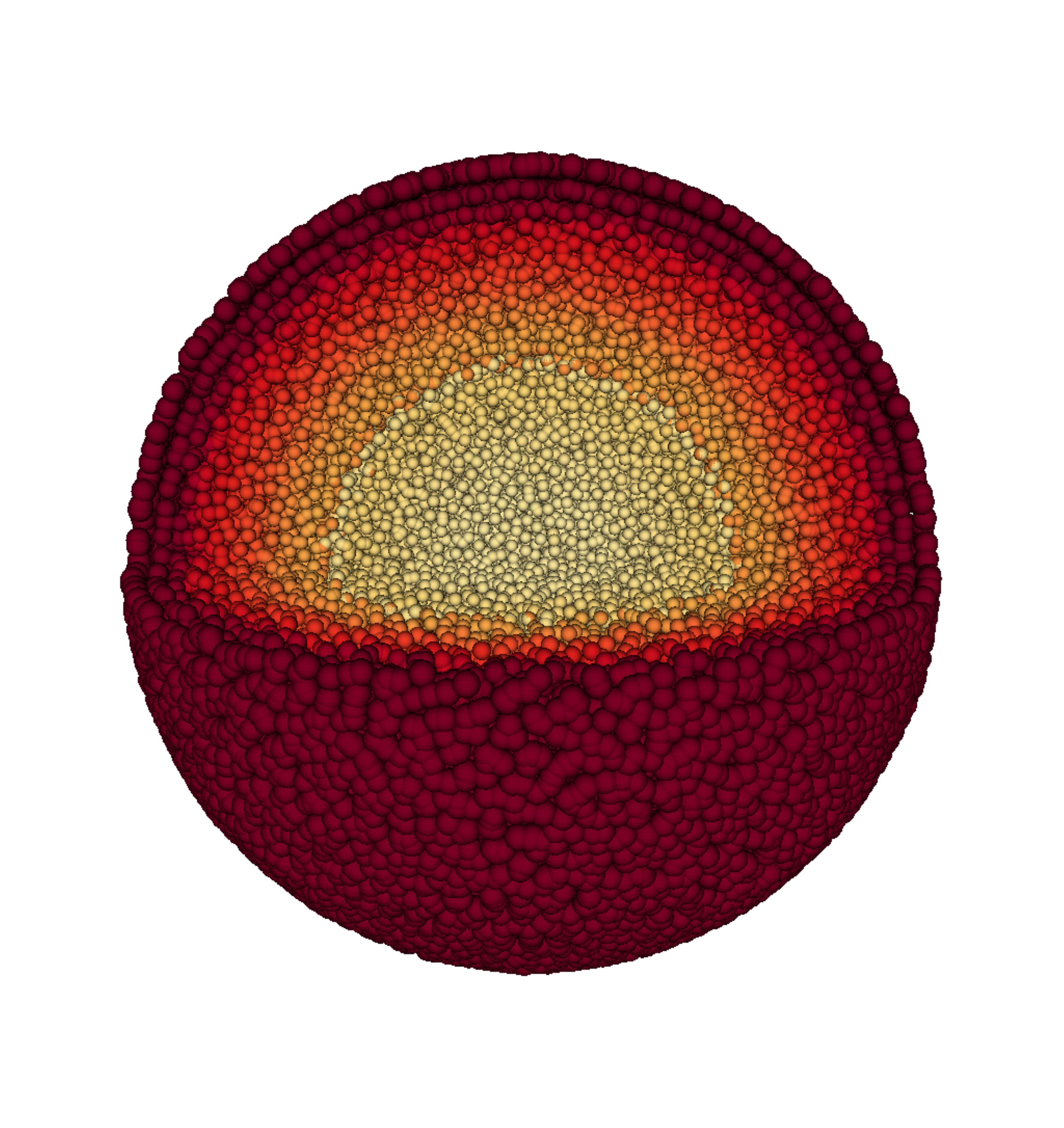}
IIc)\includegraphics[width=0.20\textwidth]{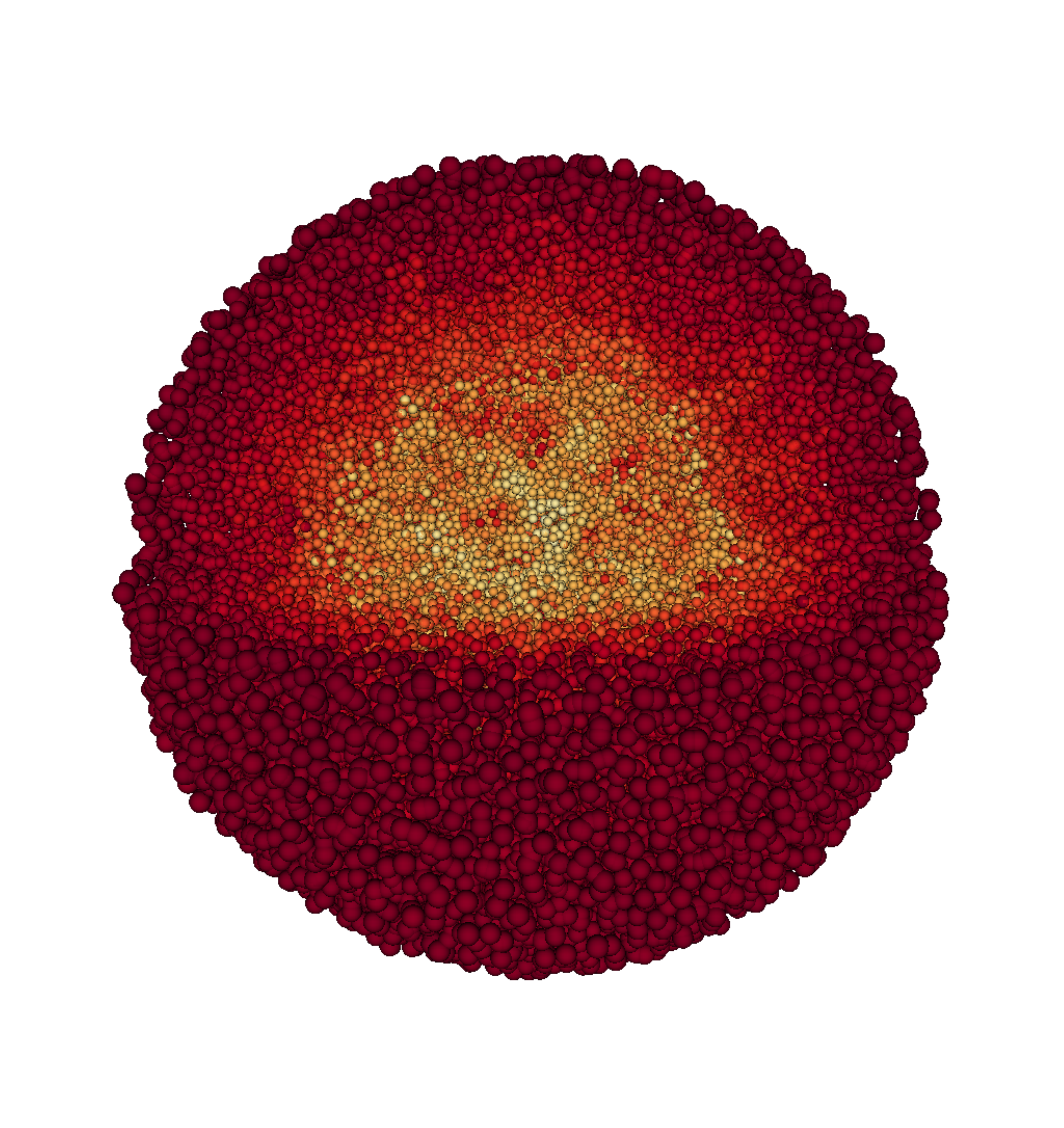}
IId)\includegraphics[width=0.20\textwidth]{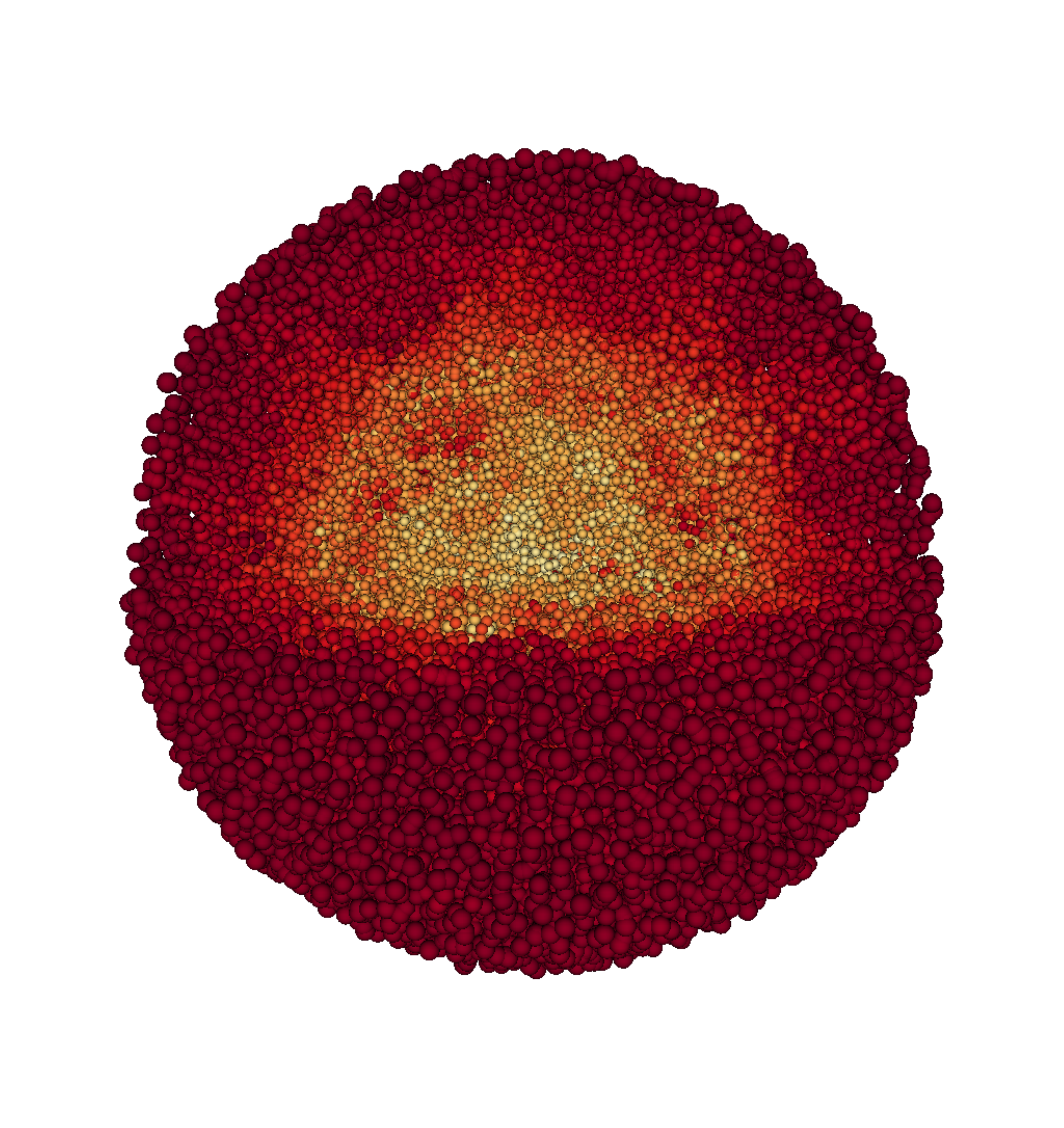}
\caption{An example distribution of $\approx100,000$ particles total with an Earth-like radial profile using a) the stretched lattice, b) the RPR+PS method, c) the 
random shell method, and d) the Monte Carlo method. The color scale indicates density. The top images (I) are the initial condition distributions with the assigned particle 
densities (not the SPH approximated densities), while the bottom images (II) are the final configurations after $t=10,000$ s ($\approx20$ sound crossing times) of 
hydrodynamical settling. For this test problem, the particle masses ranged $0.9-1.1\times$ the nominal particle mass in the refined primitives region.}
\label{fig:sphere}
\end{figure*}

\section{Results}

A particularly complicated initial condition -- from a hydrodynamical perspective -- to reproduce with any method is that of a rocky planet like the Earth. Rocky planets are built from 
concentric shells with complicated constitutive laws (equations of state) relating the pressure to the density with abrupt transitions between the shells as the material changes. For this 
reason, the equations of state can be very stiff, such that small density perturbations can result in catastrophically large pressure changes. The specific example of the Earth is primarily 
built of two very different materials; an iron-nickel core (in solid and liquid states, depending on depth), and a predominantly basaltic mantle in a mostly solid state. There is a discontinuity 
in the density function at the interface of these two materials (in fact, the Earth features several density discontinuities, one at each phase transition). As a result, any spherical 
SPH distribution that is built from a 1D profile of this kind will necessarily have a period of instability, during which the planet will oscillate as it settles into a more stable configuration. 
These oscillations offer useful diagnostics of the degrees of freedom each distribution has, and of the disparity from hydrostatic equilibrium the initial setups have.

For this paper we test the performance of the RPR+PS method against the stretched CL, random shell, and Monte Carlo methods for a two-material, single-discontinuity density 
function, using $\approx100,000$ particles. In principle any of these distributions could be used as first-pass arrangements for two-stage initial conditions generators. Here we have 
chosen to pass these arrangements through regular hydrodynamical settling, as this step is typically required of two-phase arrangements as well. The presence of the multiple spherical interfaces within 
the body make this test problem particularly difficult for non-spherically-conformal distributions, and so the results of the stretched CL can be used as an indicator of the performance of most lattice
arrangements for this particular test. The beginning and ending states of 
each of the distributions are shown in Figure \ref{fig:sphere}. The properties of these materials and the location of the interface are Earth-like, in that the core is composed of iron and 
the mantle of basalt, with an interface between the two at a depth of 3000 km (half the total radius of the object).  We use the Tillotson equation of state \citep{tillotson1962} for each of 
the two materials. Note the Tillotson does not capture the phase transitions within the core and mantle, so both materials in our test problem are single-phased. This means we have a 
single discontinuity between the core and mantle, a simpler arrangement than the actual Earth.

What is most immediately evident is the voxelization that occurs in the stretched CL arrangement as a result of the poor geometrical mapping of a CL to a 
radial density profile -- in other words, the CL arrangement is not spherically conformal. This results in the core-mantle transition straddling the CL points, leading to a 
stair-stepped interface between the two. The Monte Carlo method is also deficient in this regard as the radial positions of particles are entirely random.
In the RPR+PS and the random shell arrangements, each shell has a more-or-less spherical surface. The apparent jumbling of particles in the 
cutout of the initial RPR+PS distribution is merely the result of the random rotations applied to each shell. Seen face-on, it is clear that the RPR+PS distribution is spherically 
conformal. 

\begin{figure}[ht]
\centering
\includegraphics[width=0.40\textwidth]{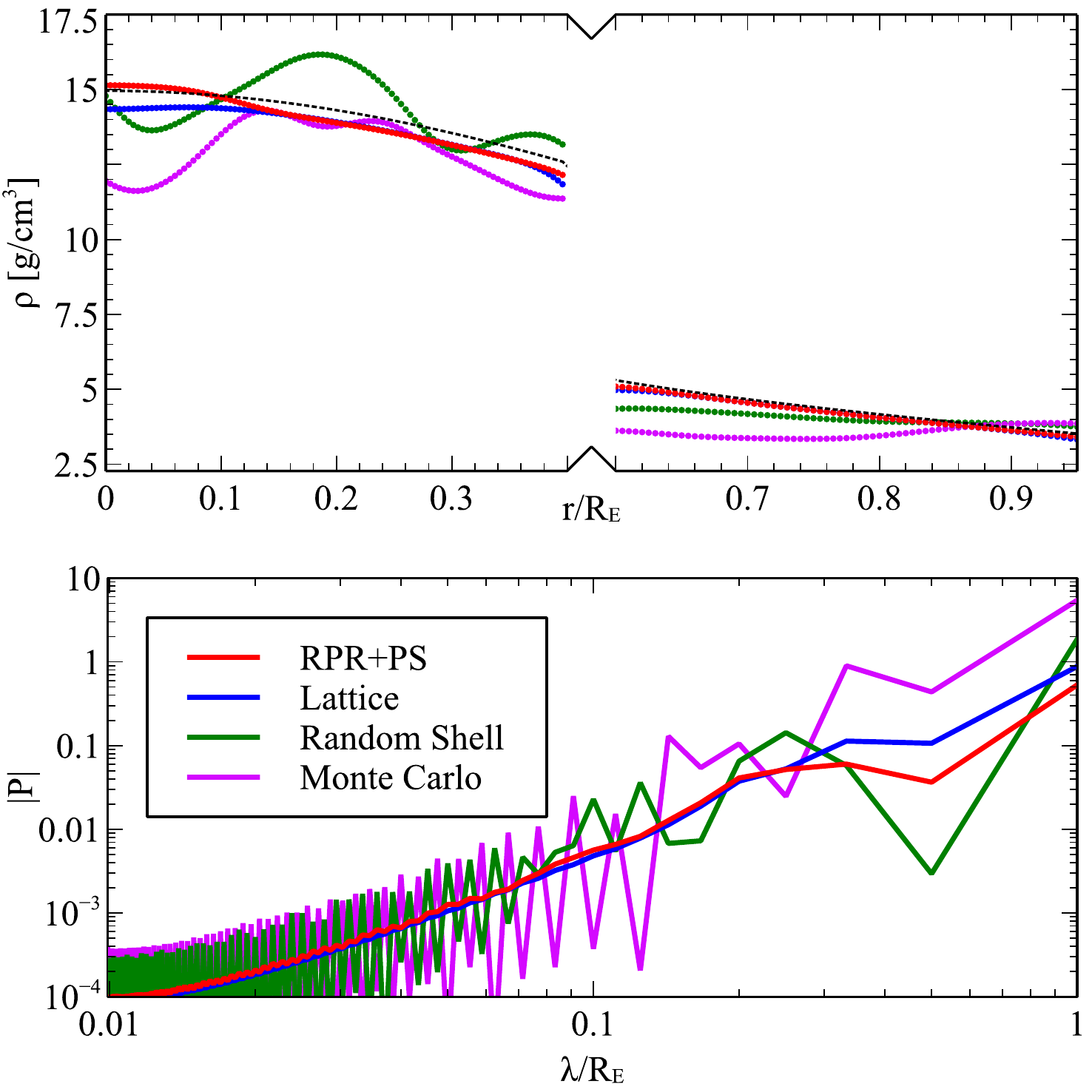}
\caption{
{\bf Top panel:} The Shepard's function approximations of the density in each of the tested distributions compared to the analytical expectation (dashed line). 
{\bf Bottom panel:} The power spectrum of the density deviations from analytical expectation.
}
\label{fig:fft}
\end{figure}

One way to measure the uniformity of a spherical initial conditions generator for SPH in the reproduction of an analytical density function is to use the SPH sum density approximation 
(equation \ref{eq:rhosph}) to find the sampled density approximation, and compare the deviations in density from the analytical function that we intended to reproduce. 
\be
  \rhosph(\bold{x}) \equiv \sum_j m_j W_i(\bold{x})
  \label{eq:rhosph}
\ee
We then sample radial profiles from these distributions' sum density approximations using a Shepard's function convolution with 100 radial bins equally spaced in $(\bold{x})$
\be
  \rho(\bold{x}) \approx \frac{\Sigma V_j\rhosph_j W_j(\bold{x})}{\Sigma V_jW_j(\bold{x})},
\ee
where $V_j$ is the particle volume ($m_j/\rho_j$), with $\rho_j$ coming from the sum density approximations, and $W_j$ is the SPH kernel function. 
Using this sampling method ensures that any deviations from the analytical expectation 
($\delta(\bold{x}) = \rho_0(\bold{x})-\rho(\bold{x})$) are due to non-unitary partitions of volume and not to the SPH kernel. Furthermore, as SPH distributions under the sum density 
approximation will err by of order a factor of two near surfaces, we constrain this portion of our analysis to the regions of our Earth model that are several particles distant from any 
surface (either between the core-mantle or at the actual outer surface) so as not to saturate the calculation of the density deviations with deviations that are an unavoidable result of 
the SPH method. Therefore we have excised the outer surface of our Earth model and also the core-mantle boundary in the calculation of $\delta(\bold{x})$. The top panel of Figure 
\ref{fig:fft} demonstrates the region of interest for this analysis. Note that the SPH approximated densities differ from the assigned particle densities in Figure \ref{fig:sphere}.

In the bottom panel of Figure \ref{fig:fft}, we show a power spectrum of the Fourier transform of $\delta(\bold{x})$ for each of our tested distributions. The two randomly seeded methods,
the Monte Carlo method and the random shell method, have the greatest power across the entire spectrum, while the RPR+PS and CL arrangements perform similarly for this metric, 
with the CL distribution having approximately twice as much power in the irreducible error that is on the scale of the entire object. 

In evolving these objects the well known surface error in the SPH mass density summation (equation \ref{eq:rhosph}) would be catastrophic for such a stiff equation of state. 
Therefore during our hydrodynamical settling we time integrate the density
based on the SPH continuity equation, making the density evolution an initial value problem which can be integrated all the way to the surface. In the final arrangements after $\approx20$ 
sound crossing times -- also shown in Figure \ref{fig:sphere} -- it is evident in the CL distribution that the grid-like nature of 
the initial setup has not entirely vanished. In addition, there is more mixing of materials across the core-mantle boundary than in the RPR+PS distribution. 

To quantify the amount of mixing we define the location below which 99\% of the iron mass resides as the maximum radius of the core, and a second location above which 99\% of the 
basalt resides as the minimum radius of the mantle. The total mass of material between these locations can then be taken as a measure of the amount of mixing across the interface. 
In the case of the CL distribution this interface has a width of $\approx160$ km, across which $\approx2.5\times10^{23}$ kg has mixed. For the random shell distribution 
$3.5\times10^{23}$ kg have mixed across a 250 km region, and for the Monte Carlo distribution $4.9\times10^{23}$ kg have mixed across a 377 km region.
For the RPR+PS distribution this interface region is 
only $15$ km wide with $7.4\times10^{22}$ kg of mixed material. The disparity here is most likely due to the conformity of the core-mantle boundary. For the CL distribution the
stair-stepping nature of this boundary results in unphysical, turbulent mixing along the non-cardinal directions. The Monte Carlo distribution should also suffer from this problem. In the 
case of the random shell distribution (where the boundary is spherically conformal) the distribution of particles is highly non-isotropic, and this too results in turbulent mixing.

\begin{figure}[ht]
\centering
\includegraphics[width=0.40\textwidth]{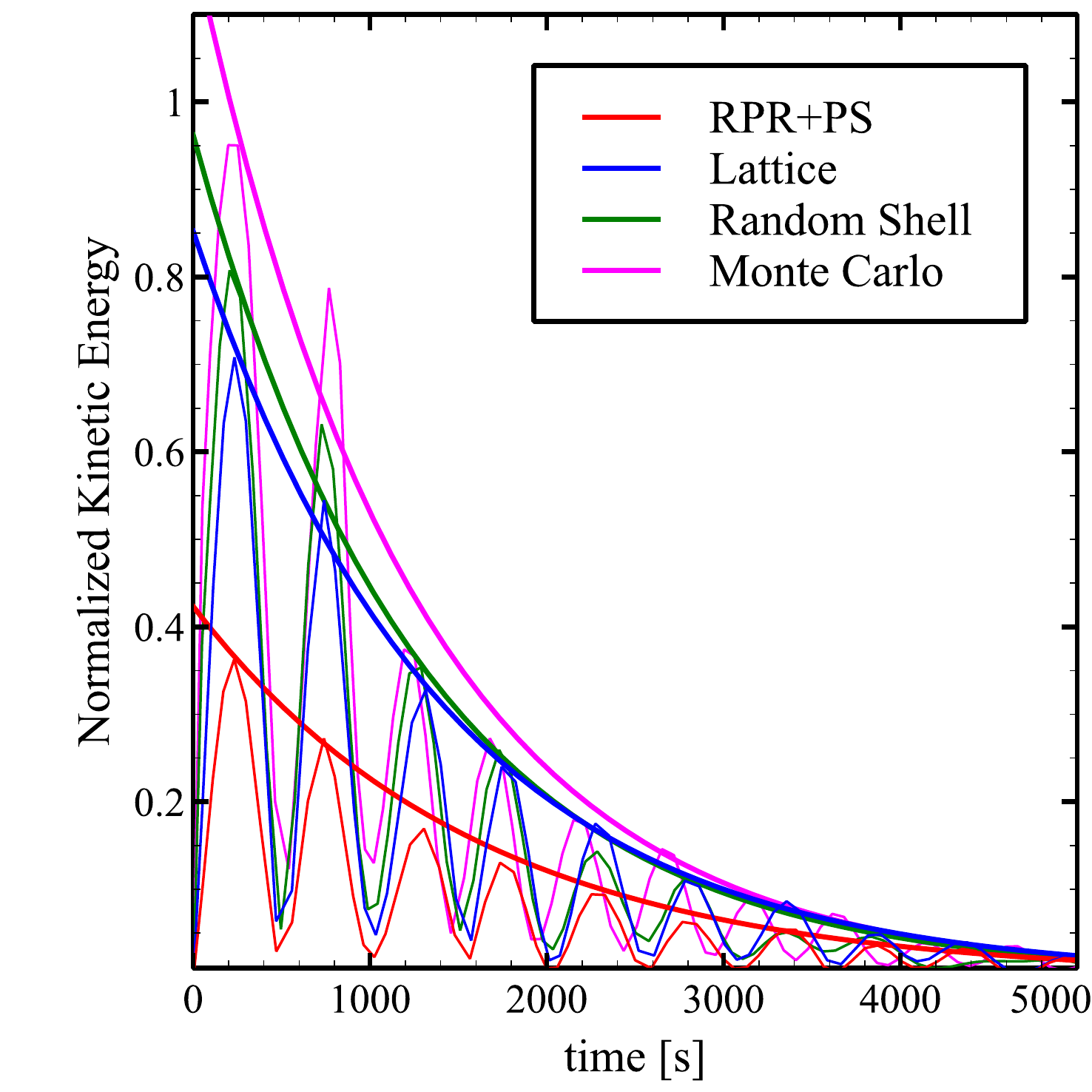}
\caption{The change in kinetic energy over time as each of the distributions undergoes gravitational settling. The RPR+PS method performs better than the other distributions on this metric 
with smaller peak amplitudes and a slightly faster oscillatory decay rate (\textit{i.e.} $e$-folding time 
$\delta t_{\rm RPR+PS}\approx1500, \delta t_{\rm CL} \approx 1700, \delta t_{\rm shell} \approx 1600,\delta t_{\rm monte carlo} \approx 1750$).}
\label{fig:settling}
\end{figure}

We plot the kinetic energy over time of the entire body in Figure \ref{fig:settling}. It is clear that the initial RPR+PS setup is nearer to equilibrium than any of the other 
setups tested here, most likely due to the interface remaining spherically symmetric as opposed to voxelized as is the case for the stretched CL. Moreover, the $e$-folding 
time for the oscillations is slightly shorter in the RPR+PS distribution, resulting in a slightly faster convergence rate. 

\begin{figure}[ht]
\centering
\includegraphics[width=0.40\textwidth]{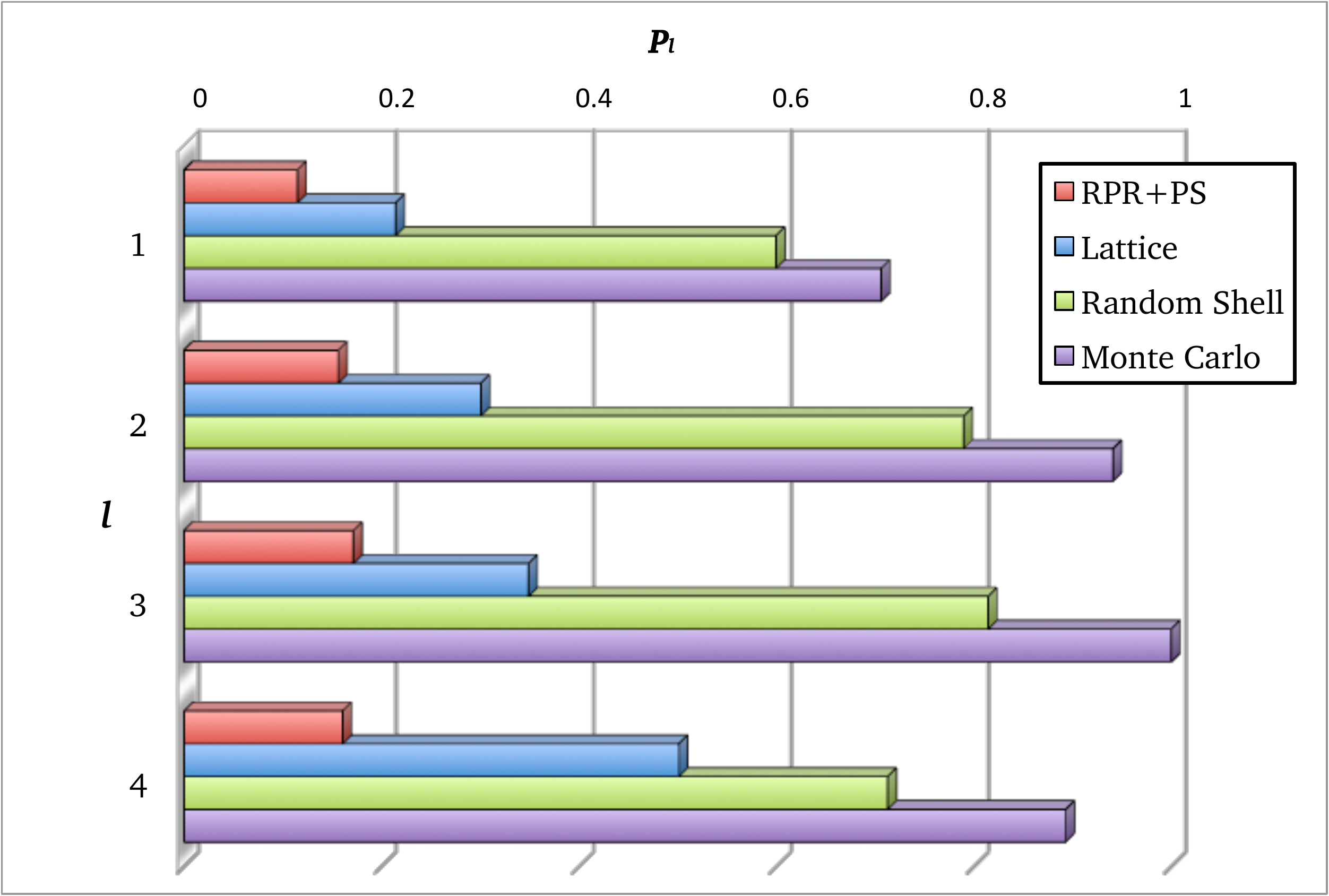}
\caption{The oscillatory power $P_l$ (summed over all $m$) normalized to $P_{3}=1$ for the monte carlo distribution on a linear scale in each of the first four harmonic modes 
(neglecting $l=0$) for each of the tested distribution methods in this paper. }
\label{fig:harmonics}
\end{figure}

Another important metric is the oscillatory power in each of the harmonic modes from these methods. To measure this, we perform a spherical harmonic decomposition of the velocity field 
at one-quarter-radius at peak velocity amplitude after two oscillatory periods. This is ample enough time for the hydrodynamics to distribute oscillatory power into the higher harmonics 
commensurate with the degrees of freedom of the initial distribution. For a perfectly spherically symmetric distribution, only the $l=0$ mode should have any oscillatory power. As is 
shown in Figure \ref{fig:harmonics}, the RPR+PS arrangement has considerably less power in the higher modes than do the others, \textit{i.e.}\ the oscillatory motion in the RPR+PS 
arrangement is much more spherically symmetric.

In the case of the RPR+PS distribution the initial configuration is conformal with spherical geometry, and so the power in non-spherical motions (higher modes) is more-or-less equal. 
By contrast in the CL distribution, the lack of spherical conformity in the initial distribution generates extra power in the higher modes. This results in deleterious non-spherical motions 
and turbulence. The random shell and Monte Carlo methods fare even worse as the random displacements away from equal volumes for all particles introduces extra noise into the 
density field.

For the random shell method the extra noise introduced into each shell tends to overwhelm any benefit derived from the spherical conformity of the shells, and the Monte Carlo method is 
simply too noisy everywhere to use as is. This is well known, and such Monte Carlo methods are best used as the first stage of a two-stage relaxation method such as the gravitational glass.
For the CL distribution the combination of the harmonic power distribution, the oscillatory magnitude and decay rate, and the turbulent mixing 
at the core-mantle boundary argues against its use for problems where sphericity is important, as in a high pressure, self-gravitating object like the Earth.  The results of the RPR+PS 
distribution, on the other hand, demonstrate its efficacy to handle these sorts of problems, mainly due to its spherical conformity by construction, and to its low-noise equipartition of area 
on surfaces.

\section{Conclusion}

The recursive primitive shape refinement in conjunction with the parameterized spiral algorithm (RPR+PS) described here offers a rapid and easily extensible way to create robust, 
spherical initial conditions in SPH for a variety of applications. Since it doesn't rely on any physics (as in two-stage generators), this method is readily adaptable as the first step in many 
SPH application scripts, obviating the need to create and store libraries of spherical particle arrangements. Alternatively, the arrangements generated by this method provide a clean initial 
basis for GG or WVT type generators as they are not plagued by random noise or aspherical geometries. Consequently, the time-to-convergence for two-stage generators ought to be 
greatly shorted, reducing computational expense.

The RPR+PS method has already been employed in the generation of initial conditions for a variety of astrophysical problems at Lawrence Livermore National Laboratories, featured in 
forthcoming papers. Examples include the construction of post-main-sequence stars in \cite{gray2016} and the construction of initial conditions for a moon-forming simulation in 
\cite{raskin2016} (using a more sophisticated equation of state than that used here), as well as various asteroid mitigation simulation tests. In each of these simulations, spherical 
conformity and radially consistent deformation are essential, and the RPR+PS method has been valuable in that regard.

The entirety of this particle distribution method was written in python as a part of the open source SPH code, \textsc{SPHERAL++}, available for download at sourceforge and can be used 
as a particle generator for any SPH code.

This work was performed under the auspices of the U.S. Department of Energy by Lawrence Livermore National Laboratory under Contract DE-AC52-07NA27344, and all tests and 
simulations were performed with computing resources provided by Lawrence Livermore National Labs, Livermore, CA. We are grateful for the geophysical consultation of Naor Movshovitz 
in the Department of Earth and Planetary Sciences at UC Santa Cruz.

\end{document}